# Drop splashing on a dry smooth surface

**Lei Xu, Wendy W. Zhang, Sidney R. Nagel\***

The James Franck Institute and Department of Physics, The University of Chicago,

5640 South Ellis Avenue, Chicago, Illinois 60637, USA.

*To whom correspondence should be addressed. Email: srnagel@uchicago.edu

**The corona splash due to the impact of a liquid drop on a smooth dry substrate is investigated with high speed photography. A striking phenomenon is observed: splashing can be completely suppressed by decreasing the pressure of the surrounding gas. The threshold pressure where a splash first occurs is measured as a function of the impact velocity and found to scale with the molecular weight of the gas and the viscosity of the liquid. Both experimental scaling relations support a model in which compressible effects in the gas are responsible for splashing in liquid solid impacts.**

What is the mechanism for the violent shattering that takes place as a liquid drop hits a smooth dry surface and splashes? How does the energy, originally distributed uniformly as kinetic energy throughout the drop, become partitioned into small regions as the liquid disintegrates into thousands of disconnected pieces? It is not surprising that the velocity of impact, the drop size and shape, or the liquid surface tension has an important effect on the mass and energy distribution of the ejected droplets [1, 2]. However, it is perhaps more difficult to imagine that the surrounding air has a significant role to play in this all-too-common occurrence. More to the point, one would hardly expect the splash to disappear if the surrounding atmosphere were removed. Nevertheless this is the case.

The elegant shapes formed during a splash have captured the attention of many photographers since the remarkable early images of Worthington showing the shapes that occur as milk or mercury hits a smooth substrate [3]. Many studies have focused on the fingering dynamics [4–7] and the effect of surface roughness [1, 2, 8]. In the present study, we focus only on a drop hitting a smooth substrate. The top row of Figure 1 shows four frames from a movie of an alcohol drop hitting a dry glass slide in a background of air at atmospheric pressure. The drop, after impact, spreads and creates a corona with a thickened rim which first develops undulations along the rim and then breaks up due to surface tension. During this process, the thin sheet comprising the corona surface retracts and rips into pieces. These images are reminiscent of the corona caused by a drop hitting a thin layer of fluid photographed by Edgerton and his colleagues [9]. However, in our case we have made sure that the slide is completely dry prior to impact. Our images illustrate an important puzzle: why do we see a corona form at all? At the substrate surface the liquid momentum points horizontally outward. Without a layer of fluid to push against (such as in the photographs of Edgerton), how does the expanding layer gain any momentum component in the vertical direction?

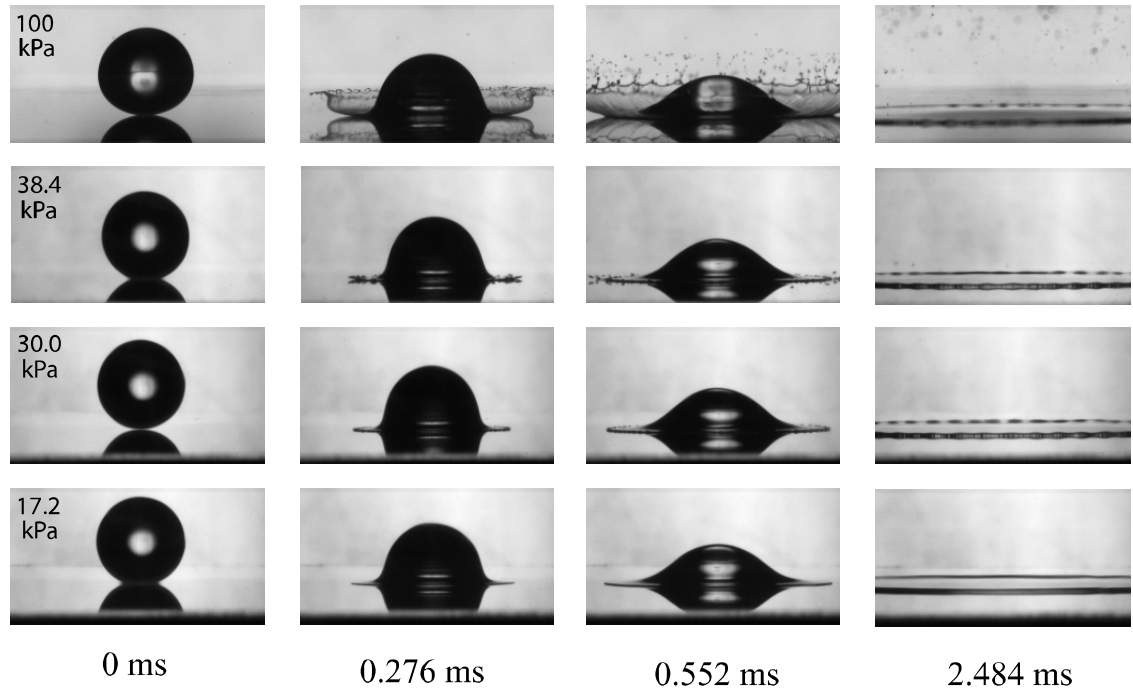

0 ms          0.276 ms          0.552 ms          2.484 ms

**Fig. 1.** Photographs of a liquid drop hitting a smooth dry substrate. A 3.4 ± 0.1 mm diameter alcohol drop hits a smooth glass substrate at impact velocity $V_0 = 3.74 \pm 0.02$ m/s in the presence of different background pressures of air. Each row shows the drop at four times. The first frame shows the drop just as it is about to hit the substrate. The next three frames in each row show the evolution of the drop at 0.276 ms, at 0.552 ms and at 2.484 ms after impact. In the top row, with the air at 100 kPa (atmospheric pressure), the drop splashes. In the second row, with the air just slightly above the threshold pressure, $P_T = 38.4$ kPa, the drop emits only a few droplets. In the third row, at a pressure of 30.0 kPa, no droplets are emitted and no splashing occurs. However, there is an undulation in the thickness of the rim. In the fourth row, taken at 17.2 kPa, there is no splashing and no apparent undulation in the rim of the drop.

Our experiment is straightforward: Reproducible drops of diameter D = 3.4±0.1 mm were released from rest at different heights above a glass microscope slide laid horizontally inside a transparent vacuum chamber. The pressure, P, could be varied between 1 kPa and 100 kPa (atmospheric pressure) and the height of the nozzle above the substrate could be varied between 0.2 m and 3.0 m. The subsequent splash was recorded by a Phantom V7 high-speed video camera at a frame rate of 47,000 fps. The impact speed of the drop was also determined from these movies. Because the drop shape oscillates after it leaves the nozzle, we adjusted the height carefully so that, in all the measurements reported below, the profile of the drop was nearly circular at the instant that it made contact with the slide. Also, in order to avoid contamination of the glass due to the possible residue left by previous drops, we replaced the substrate with a fresh slide after every measurement. We have used three different liquids (methanol, ethanol and 2-propanol) for the drop and four different gases (helium, air, krypton and $SF_6$) for the surrounding atmosphere. The liquids that we chose all wet the substrate so that there is no subsequent retraction and rebound of the drop [10].

The rows of Fig. 1 show images of the splash at different background air pressures for a drop of ethanol hitting the substrate at a velocity $V_0$ = 3.74 ± 0.02 m/s. Surprisingly, as the pressure is lowered, fewer droplets are ejected from the surface; below P = 30 kPa no droplets emerge at all after impact. We are able to determine the threshold pressure at which splashing occurs, $P_T$ [11], as a function of impact velocity, $V_0$. In Fig. 2a, we show $P_T$ versus $V_0$. As expected over most of the range, $P_T$ decreases as the impact velocity is raised. However, there is one region below V*, where this is not true and the curve is non-monotonic. In this low velocity regime, splashing is doubly re-entrant. As we will show, this effect appears with other liquids and other gases.

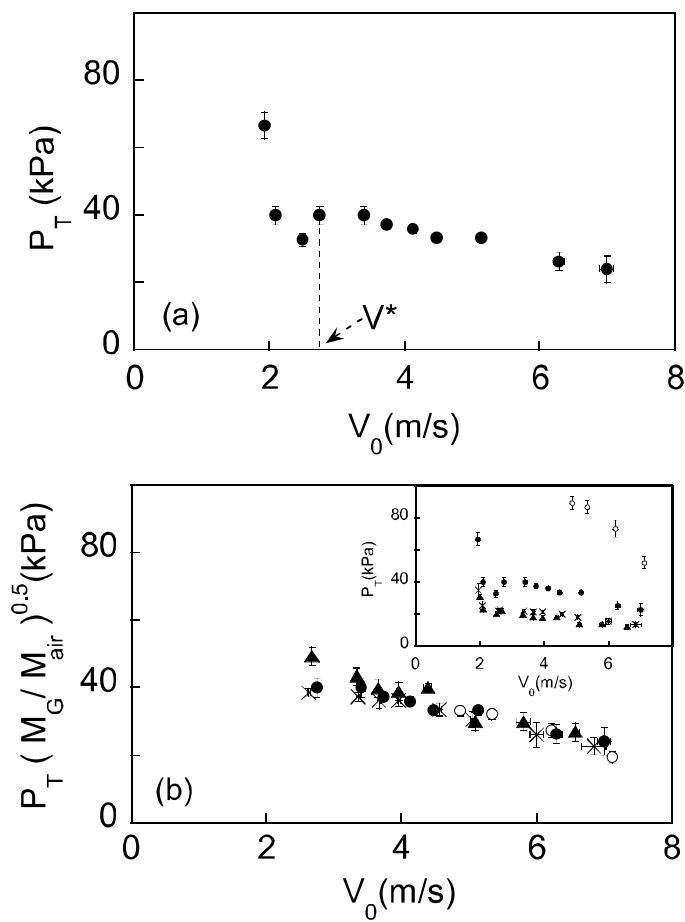

**Fig. 2.** Threshold pressure versus impact velocity. (**a**), $P_T$ is plotted versus $V_0$ in a background atmosphere of air. The data is non-monotonic in the region $V_0 < V^*$. The lower end of the error bars gives the pressure where droplets just begin to be discharged disconnectedly from the rim of the expanding liquid layer; the upper end of the error bars gives the pressure where fully developed splashes emanate uniformly along the entire rim. The midpoint is defined as the threshold pressure. (**b**), Inset shows $P_T$ versus $V_0$ for four gases: He(O), air(●), Kr(X) and $SF_6$(▲). The main panel shows the scaled threshold pressure, $P_T(M_G/M_{air})^{0.5}$, versus the impact velocity, $V_0$, in the region $V_0 > V^*$, for the four gases shown in the inset.

Clearly the pressure of the gas is essential for determining whether or not the drop will splash. However it is not obvious what physical property of the gas is important. We note that the dynamic viscosity of the air does not vary with pressure until the mean free path of the molecules is the size of the geometric length scales of the system. We are well above that regime in these experiments. We also measured [12] that the surface tension of the liquid does not vary with pressure in our experimental regime. In order to understand its role better, we have varied the composition of the gas. The inset to Fig. 2b shows the threshold pressure versus impact velocity for four different gases; the values of $P_T$ are displaced from each other but the trends in the data have the same qualitative shape. We note that the four gases used have similar viscosities (varying only from 15.3 µPa s for $SF_6$ to 25.6 µPa s for Kr) [13] but have very different molecular weights, $M_G$ ($M_{He} = 4$, $M_{air} = 29$, $M_{Kr} = 83.8$, and $M_{SF6} = 146$ Daltons) [13]. We have tried to scale the different curves on top of one another and found that the best data collapse, in the region with impact velocities greater than V*, is obtained by plotting $(M_G/M_{air})^{0.5} P_T$ versus $V_0$. The result is shown in the main panel of Fig. 2b.

Our analysis concentrates entirely on the regime with $V_0 > V*$. We consider two contributions to the stress on the expanding liquid layer after impact: the first, $\Sigma_G$, is due to the restraining pressure of the gas on the spreading liquid, which acts to destabilize the advancing front and deflect it upward; the second, $\Sigma_L$, is due to the surface tension of the liquid, which favours keeping the liquid layer intact after impact. When the two stresses become comparable, we expect the spreading liquid to become unstable and to break up into droplets.

On impact, the drop spreads out suddenly and rapidly. An estimate of $\Sigma_G$ should therefore include the effects of the shock wave that the liquid creates in the air. We apply the water-hammer equation [8] to the gas phase [14] which states that the stress is proportional to the gas density, $\rho_G$, the speed of sound in the gas, $C_G$, and the expanding velocity of the liquid layer on the substrate, $V_e$:

$$\Sigma_G \sim \rho_G \cdot C_G \cdot V_e \sim \frac{PM_G}{k_B T} \cdot \sqrt{\frac{\gamma k_B T}{M_G}} \cdot \sqrt{\frac{RV_0}{2t}} \quad .$$  (1)

Here $\gamma$ is the adiabatic constant of the gas, T is the temperature, $k_B$ is Boltzmann's constant, R is the initial radius of the drop and t is the time measured from the instant of impact.

In order to estimate $\Sigma_L$, we consider the surface tension pressure near the front of the advancing liquid. This is given by the surface tension coefficient, $\sigma$, divided by the thickness of the layer, d. The thickness d is assumed to be the boundary layer thickness which is controlled by the diffusion of vorticity from the solid substrate [15]. Thus:

$$\Sigma_L = \sigma \ / \ d = \sigma \ / \ \sqrt{v_L t}$$  (2)

where $v_L$ is the kinematic viscosity of the liquid. These estimates imply:

$$\frac{\Sigma_G}{\Sigma_L} = \sqrt{\gamma M_G} \, P \cdot \sqrt{\frac{RV_0}{2k_B T}} \cdot \frac{\sqrt{v_L}}{\sigma}$$  (3)

which is independent of time. When the two stresses are comparable, the expanding liquid rim is slowly destabilized and deflected upwards for an extended period of time, finally resulting in the ejection of droplets. This equation predicts another non-intuitive result: a more viscous liquid splashes more easily than a less viscous one. That is, the threshold pressure should decrease if the liquid viscosity is raised. To test this prediction, we have studied splashes from three different alcohols (methanol, ethanol and 2-propanol) that have essentially the same density and surface tension but different kinematic viscosity ($v_{meth} =$ $0.68$, $\nu_{eth} = 1.36$, and $\nu_{2\text{-prop}} = 2.60$ $\mu$Pa s m$^3$/kg) [13]. The results for $P_T$ versus $V_0$ are shown in the inset to Fig. 3. Indeed, it is the case that 2-propanol, the liquid with the largest viscosity, has the lowest threshold pressure. The main panel of Fig. 3 shows $P_T$ ($\nu_L$ /$\nu_{eth}$)$^{0.5}$ versus $V_0$ for all the liquids studied in the regime $V_0 > V^*$. There is a good collapse of the data. In Fig. 4, we show the ratio: $\Sigma_G/\Sigma_L$ at the threshold pressure for all our data with $V_0 > V^*$. The ratio is approximately constant, independent of impact velocity, with $\Sigma_G/\Sigma_L \sim 0.45$. This indicates that $\Sigma_G$ and $\Sigma_L$ are comparable at the threshold pressure, as we expected.

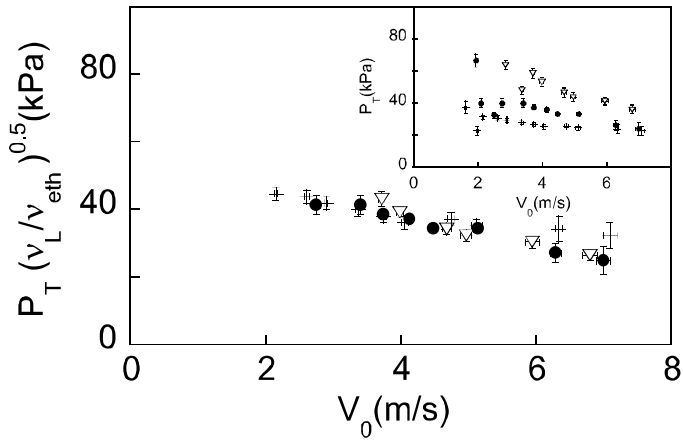

**Fig. 3.** Effect of liquid viscosity. Inset shows $P_T$ versus $V_0$ for three liquids: methanol ($\triangledown$), ethanol ($\bullet$) and 2-propanol (+), in a background atmosphere of air. Main panel shows the scaled threshold pressure, $P_T$ ($\nu_L/\nu_{eth}$)$^{0.5}$, versus the impact velocity, $V_0$, in the region $V_0 >$ $V^*$, for the three liquids shown in the inset.

We have shown that, surprisingly, the presence of a surrounding gas is essential for splashing to occur on a dry flat substrate. Moreover, it provides a means for creating the corona with a vertical component of momentum which would be difficult to produce without gas being present. Several puzzles remain. Although we have made an estimate, which concurs with the experimental data, for where splashing should occur if the impact velocity is sufficiently large, we have no similar estimate for what should happen in the low-velocity regime. Indeed, we do not yet know why there is a separate regime at small $V_0$. Likewise, we suspect that there may be other regimes, for example when the liquid viscosity becomes large or when the impact velocity becomes comparable to the sound speed in the gas.

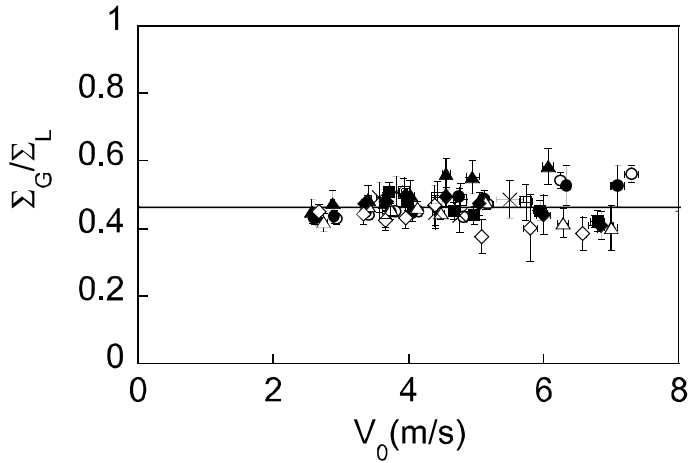

**Fig. 4.** Ratio at threshold of $\Sigma_G$, the destabilizing stress due to the gas, to $\Sigma_L$, the stabilizing stress due to surface tension. $\Sigma_G / \Sigma_L$ is plotted versus $V_0$, in the region $V_0 > V^*$, for all the liquids and gases studied.

The discovery that the surrounding pressure and gas composition can influence the occurrence of splashes, should have important technological ramifications in the many situations where splashing is involved such as in combustion of liquid fuels [16], spray drying [17], ink-jet printing [18], and industrial washing. For example in the case of surface coating, where splashing causes problems, we can either pull a vacuum or simply vary the composition of the gas to one with a low molecular weight. In other cases, where splashing is desired, we can do just the opposite. This provides a technique to control splashing precisely.

We wish to thank Christophe Clanet, Itai Cohen, Haim Diamant, Christophe Josserand, Stephan Zaleski, and Ling-Nan Zou for helpful discussions. This work was supported by NSF 0352777 and NSF MRSEC DMR-0213745.